\documentclass[aps,prd,nofootinbib,
twocolumn,
superscriptaddress,floatfix]{revtex4}
 
\usepackage{amsmath,amssymb}
\usepackage{mathrsfs}
\usepackage[tight,footnotesize]{subfigure}
\usepackage[dvipsnames]{xcolor}
\usepackage{float}
\usepackage{wrapfig}
\usepackage{hyperref}
\usepackage{mathtools}
\usepackage[utf8]{inputenc}
\usepackage{graphicx}
\usepackage{physics}

\usepackage{bm}
\usepackage{braket}
\usepackage{listings}
\usepackage{cases}
\usepackage{comment}
\usepackage{soul}
\usepackage{cancel}
\usepackage{xspace}
\usepackage{acronym}
\usepackage{siunitx}

\newcommand{\munu}{_{\mu\nu}}
\newcommand{\MUNU}{^{\mu\nu}}

\newcommand{\pt}[1]{\partial_{#1}}

\renewcommand{\ij}{{}_{ij}} 
\renewcommand{\IJ}{{}^{ij}}

\newcommand{\Df}{\mathcal{D}}

\newcommand{\gm}{\gamma}

\newcommand{\bea}{\begin{eqnarray}}
\newcommand{\eea}{\end{eqnarray}}

\newcommand{\bae}[1]{\begin{align} #1 \end{align}}
\newcommand{\beae}[1]{\begin{equation}\begin{aligned} #1 \end{aligned}\end{equation}}
\newcommand{\bege}[1]{\begin{equation}\begin{gathered} #1 \end{gathered}\end{equation}}

\newcommand{\bfe}[4]{
\begin{figure} 
	\centering
	\includegraphics[#1]{#2}
	\caption{#3}
	\label{#4}
\end{figure}}

\newcommand{\ini}{\mathrm{ini}}
\newcommand{\diag}{\mathrm{diag}}

\newcommand{\calD}{\mathcal{D}}

\newcommand{\calH}{\mathcal{H}}

\newcommand{\calR}{\mathcal{R}}

\newcommand{\be}{\begin{equation}}
\newcommand{\beqn}{\begin{eqnarray}}
\newcommand{\eeq}{\end{equation}}
\newcommand{\eeqn}{\end{eqnarray}}

\usepackage{xcolor}

\begin{document}
\title{
Spacetime Evolution 
Induced by a Massive Cosmic String
in Numerical Relativity
}

\author{Yoshihiro Kishimoto}
\affiliation{Department of Physics, Institute of Science Tokyo, 2-12-1 Ookayama, Meguro-ku, Tokyo 152-8551, Japan}

\author{Cristian Joana}
\affiliation{Center for Theoretical Physics, School of Physics and Optoelectronic Engineering,\\ Hainan University, Haikou 570228, China}
\affiliation{International Centre for Theoretical Physics Asia-Pacific,
University of Chinese Academy of Sciences, 100190 Beijing, China}

\author{Teruaki Suyama}
\affiliation{Department of Physics, Institute of Science Tokyo, 2-12-1 Ookayama, Meguro-ku, Tokyo 152-8551, Japan}

\date{\today}

\begin{abstract}
Cosmic strings are one-dimensional topological defects formed during phase transitions in the early universe.
``Massive'' cosmic strings possess a sufficiently large tension to close their transverse space, forming a conical singularity outside the string, as a static gravitational effect.
In a dynamical context, such strings are expected to drive topological inflation.
In this work, we investigate the spacetime evolution induced by the formation of a massive cosmic string in a radiation-dominated cosmological background, imposing cylindrical symmetry.
Using full numerical relativity simulations, we observe that the string core undergoes topological inflation and that its boundary moves inwards superluminally, which causally disconnects the core from the outer universe.
Between the core and the asymptotic regions, a nearly matter-dominated decelerating region originates sourced by the oscillating string fields.
This region is bounded by a local  maximum (``belly'') and local minimum (``throat'')  of the circumferential radius, while the throat recedes outwards at about 10\% the speed of light.
Furthermore, a gravitational wavefront is emitted from the string, characterizing the light cone of the string formation.
Even behind the wavefront, the exterior region no longer exhibits a ``conical'' structure.
Based on the simulation results, we construct a Penrose diagram, reflecting the causal structure of the entire spacetime.
\end{abstract}

\maketitle

\section{Introduction}\label{sec:intro}
Cosmic strings are one-dimensional topological defects which can form during phase transitions with spontaneous symmetry breaking in the early universe~\cite{vilenkin2000Cosmic, nielsen1973Vortex, hindmarsh1995Cosmic}.
In field theory, cosmic strings are solitons realized when the vacuum manifold has a non-trivial fundamental group, and their formation is predicted in many theories beyond the Standard Model~\cite{jeannerot2003How}.
The simplest cosmic string model is the global string, appearing in the theory of a complex scalar field involving a spontaneous global $\mathrm{U}(1)$ symmetry breaking.
However, the energy per unit length of a global string diverges.
The next simplest model, called the local string, is obtained by gauging the $\mathrm{U}(1)$ symmetry, known as the Abelian-Higgs model.
For a local string, the core consists of the magnetic flux and the vacuum energy, both of which decay exponentially toward the true vacuum state away from the core.
The tension $\mu$ of a local string scales as $\eta^2$, where $\eta$ is the symmetry-breaking scale.

The spacetime geometry induced by a straight and static cosmic string is locally flat outside the core, but globally conical~\cite{vilenkin1981Gravitational, gott1985Gravitational, hiscock1985Exact}.
The deficit angle of the conical space is $8\pi G\mu$ to leading order in $G\mu$~\cite{vilenkin2000Cosmic}.
The time-dependent gravitational effects induced by the formation of a straight string have also been solved analytically within the framework of cosmological perturbation theory using the wire approximation of the string~\cite{magueijo1992Inborn, lake2012Evolution}.
The formation effects propagate separately at the speed of light and the speed of sound, approaching a self-similar evolution in which the standard conical structure emerges only deep inside the sound horizon.

The conical description breaks down for string tensions close to the Planck scale ($8\pi G\mu\gtrsim 2\pi$).
The static gravity of such a string causes the transverse space to close and generate a conical singularity outside the string~\cite{gott1985Gravitational}.
While regularizing the additional conical singularity has been studied~\cite{blanco-pillado2014Supermassive}, an alternative approach is provided by topological inflation~\cite{linde1994Topological, vilenkin1994Topological}.
Topological inflation is inflationary expansion inside the core of topological defects, driven by the false-vacuum energy trapped within the core.
It occurs when the Hubble length associates to this vacuum energy becomes comparable to, or smaller than, the defect's core size. 
For a cosmic string, this condition is equivalent to a tension close to the Planck scale, $G\mu\sim1$.
This value is close to the tension at which the static gravity causes the transverse space to close, forming a conical singularity. We refer to such a string as a ``massive string''.

The realization and threshold of topological inflation have been investigated for monopoles~\cite{sakai1996Dynamicsb, cho1997Spacetime, sakai2006Universe}, and strings~\cite{cho1998Inflation}.
In spherically symmetric systems, such as vacuum bubbles, the resulting spacetime structure is well established, where a black hole forms and separates the inflating core from the outer region~\cite{blau1987Dynamicsb, farhi1990It}.
However, for cosmic strings, while Cho has numerically demonstrated topological inflation of a string~\cite{cho1998Inflation}, the causal structure and the fate of the spacetime remain unclear.
Moreover, describing a massive string whose gravitational field is initially localized requires a non-vacuum cosmological background.
Therefore, we focus on the evolution of spacetime induced by the formation of a massive string in a Friedmann–Lemaître–Robertson–Walker (FLRW) universe. 

Tracking topological inflation and its backreaction requires solving the full set of coupled field equations for the scalar field, gauge field, perfect fluid, and gravitational field.
The system is fully non-linear and lies beyond the scope of perturbative analysis.
To simulate this system, we employ numerical relativity, which is a well-established framework that enables us to solve the non-linear dynamics and to extract its causal structure.
We restrict our analysis to cylindrical symmetry, to make the system computationally tractable.

This paper is organized as follows.
In Section ~\ref{sec:string}, we introduce the details of the physical situation and the full action of our system and formulate a cylindrically symmetric ansatz.
Section ~\ref{sec:setup} details our numerical methodology and simulation setup.
In Section ~\ref{sec:result}, we present our main results regarding the evolution of the spacetime.
Finally, the deduced causal structure and further discussion are provided in Section ~\ref{sec:discussion}.
Technical details on the numerical implementation, the construction of the initial string profile, the slicing condition, and code validation are collected in Appendix. 
Throughout this paper, we adopt the signature of the metric as $(-,+,+,+)$ and work in the natural units $c=\epsilon_0=\hbar=1$,  keeping Newton’s constant $G$ explicit.

\section{Straight cosmic string with gravity}\label{sec:string}
Our focus is the evolution of a cylindrically symmetric spacetime induced by a massive local string within an unclosed and infinitely extended space, in which the gravitational effects of the string are initially localized.
However, it is known that the static gravity of the massive local string causes the perpendicular space to close~\cite{gott1985Gravitational}.
To avoid this and obtain localized initial data, the spacetime cannot be asymptotically Minkowski.
Moreover, cosmic strings naturally form during a second-order phase transition in an expanding universe, rather than in flat spacetime. To consistently satisfy these requirements, we adopt a flat radiation-dominated FLRW universe as the asymptotics of our system.
Note that ``asymptotic'' here refers to the region far from the symmetry axis since the region at infinity in the axial direction does not constitute an asymptotic region in a cylindrically symmetric spacetime.
At the same time, our system can be physically interpreted as a scenario where the massive string is formed after the corresponding second-order phase transition.
The field oscillations after the transition are assumed to have been converted into a radiation via quantum processes, producing the flat radiation-dominated FLRW background. 
Such reheating-like processes require additional fields coupling to the string fields; modeling them is beyond our scope, and we neglect them for simplicity.

We now introduce the model describing the system.
The action considered here is
\bae{S=\int d^4x \sqrt{-g}\qty(\mathcal{L}_{\mathrm{EH}} + \mathcal{L}_{\mathrm{AH}} + \mathcal{L}_{\mathrm{fluid}}),
\label{eq:action}}
where $\mathcal{L}_{\mathrm{EH}} = \calR/16\pi G$ is the Einstein-Hilbert Lagrangian and the remaining terms
describe the matter sector.
The Abelian-Higgs sector $\mathcal{L}_{\mathrm{AH}}$ contains a complex scalar field $\phi$ with a Mexican hat potential and a U(1) gauge field $A_{\mu}$:
\bege{\mathcal
    {L}_{\rm AH} = -\Df_{\mu}\phi(\Df^{\mu}\phi)^* - V(\phi) - \frac{1}{4}F\MUNU F\munu \\
    V(\phi) = \frac{\lambda}{4}\qty(\abs{\phi}^2-\eta^2)^2
    \label{eq:AH_lagrangian},}
where $\Df_{\mu}=\nabla_{\mu}-ieA_{\mu}$ is the U(1) covariant derivative and $e$ is the U(1) charge of $\phi$, and $F\munu=\pt{\mu}A_{\nu}-\pt{\nu}A_{\mu}$ is the U(1) gauge field strength.

In cylindrical coordinates $(t,r,\theta,z)$,
we adopt the following ansatz for a cylindrically symmetric string, 
\begin{align}
    \phi=\eta f(t,r) e^{in\theta},\quad 
    A_{\mu} = \frac{n}{e} g(t,r)\delta^{\theta}_{\mu},
\label{eq:cylind_sol}
\end{align}
where $n$ is the winding number of the string, and $f$ and $g$ are dimensionless variables.
For nonzero winding, regularity requires the amplitudes to vanish at the core of the symmetry axis, $f(t, r\!=\!0)=g(t, r\!=\!0)=0$.
Away from the core, while the potential favors $f\rightarrow1$, the gauge field regulates the angular variation of the scalar phase through the covariant derivative $\calD_\theta\phi\rightarrow in(1-g)\eta e^{in\theta}$.
Far from the core, where $g\rightarrow1$, the gauge contribution cancels the phase derivative, preventing the winding from producing a long-range scalar gradient energy. 
We then impose the boundary conditions
\bege{f(t,0)=0,\quad \lim_{r\rightarrow\infty}f(t,r)=1 \\
g(t,0)=0,\quad \lim_{r\rightarrow\infty}g(t,r)=1
\label{eq:NO_boundary}}
to realize a cylindrically localized cosmic string.
For a static string in Minkowski spacetime, this ansatz yields the Nielsen--Olesen solution~\cite{nielsen1973Vortex}.

To render the variables dimensionless, we introduce new variables via the relations $x^{\mu}=(e\eta)^{-1} x_{\mathrm{new}}^{\mu}$, $\phi=\eta \phi_{\mathrm{new}}$, and $A_{\mu}=e\eta A^{\mathrm{new}}_{\mu}$ for $\mu\neq\theta$, while $\theta=\theta_{\mathrm{new}}$ and $A_{\theta}=A^{\mathrm{new}}_{\theta}$, since the angular coordinate $\theta$ is dimensionless by definition.
For simplicity of notation, we omit the ``new'' labels and retain the original symbols for these dimensionless variables.
Under this replacement, the Lagrangian scales as
\begin{equation}
    \mathcal{L}_{\mathrm{EH}} + \mathcal{L}_{\mathrm{AH}} \rightarrow e^2\eta^2(\mathcal{L}_{\mathrm{EH}} + \eta^2\mathcal{L}_{\mathrm{AH}}) ~.
\end{equation}
Moreover, with the ansatz~\eqref{eq:cylind_sol}, the scaled Abelian-Higgs Lagrangian becomes
\beae{\mathcal{L}_{\mathrm{AH}} =& -\pt{\mu}f\pt{}^{\mu}f - n^2f^2\qty(g^{\theta\theta} - 2g^{\theta\mu}\tilde{A}_{\mu}+\tilde{A}_{\mu}\tilde{A}^{\mu} ) \\
&-\frac{\beta}{2}\qty(f^2-1)^2 - \frac{1}{4}\tilde{F}\munu\tilde{F}\MUNU
\label{eq:rescaled_AH_lagrangian},}
where $\tilde{A}_{\mu}=g\delta_{\mu}^{\theta}$, and $\tilde{F}\munu=\pt{\mu}\tilde{A}_{\nu}-\pt{\nu}\tilde{A}_{\mu}$ is its corresponding field strength tensor.
The scaled Abelian-Higgs Lagrangian $\mathcal{L}_{\mathrm{AH}}$ has two dimensionless parameters, $\beta=\lambda/2e^2$ and $n$, while $\eta^2$ identifies the effective scale of the energy-momentum tensor.
Consequently, the condition for topological inflation is determined dominantly by $G\eta^2$, as shown in~\cite{cho1998Inflation}.
It is worth noting that the physical scale of the false-vacuum energy driving topological inflation is $\lambda\eta^4$ ($=2e^2\beta\eta^4$).
Thus, $G\eta^2\sim1$ does not necessarily imply a Planckian energy density, provided that $\lambda$ is small.

The perfect fluid sector $\mathcal{L}_{\mathrm{fluid}}$ is introduced to realize an asymptotically FLRW universe.
The energy-momentum tensor is given by
\begin{align}
    T^{\mathrm{fluid}}\munu = (\rho_{\mathrm{f}}+p_{\mathrm{f}})u_{\mu}u_{\nu} + p_{\mathrm{f}}\,g\munu
    \label{eq:PF_emtensor},
\end{align}
where $\rho_{\mathrm{f}}$ and $p_{\mathrm{f}}$ are the proper energy density and pressure of the fluid, respectively, and $u^{\mu}$ is the fluid four-velocity.
We assume a barotropic equation of state $p_{\mathrm{f}}=w\rho_{\mathrm{f}}$ with the equation of state parameter $w=1/3$, which corresponds to radiation.
This perfect fluid couples to the Abelian-Higgs sector solely through gravity.
Note also that the coordinate variables are scaled by $e\eta$ and are dimensionless.
Thus, under the scaling $T^{\mathrm{fluid}}\munu\rightarrow e^2\eta^2T^{\mathrm{fluid}}\munu$, $\rho_{\mathrm{f}}$ and $p_{\mathrm{f}}$ carry the dimensions of mass squared.
The evolution equations and the energy-momentum tensor of the system~\eqref{eq:action} are given in Sec.~\ref{sec:eqs}.

\section{Numerical Setup}\label{sec:setup}

\subsection{BSSN Formalism}
We solve the Einstein equations using the Baumgarte--Shapiro--Shibata--Nakamura (BSSN) formalism~\cite{baumgarte1998numerical, shibata1995Evolution} of numerical relativisty.
Its formulation in curvilinear coordinates has been developed for spherical~\cite{baumgarte2013Numerical} and cylindrical coordinates~\cite{alcoforado2026Baumgarte}.
Here, we apply this formalism to a cylindrically symmetric spacetime, which corresponds to adding a translational symmetry in the $z$-direction to the formulation in~\cite{alcoforado2026Baumgarte}.
In this formalism, the spacetime metric is ADM decomposed, and the line element reads
\begin{align}
    ds^2 = -\alpha^2 dt^2 + e^{4\psi}\bar{\gm}\ij(\beta^i dt+dx^i)(\beta^j dt+dx^j)
\end{align}
where $\alpha$ is the lapse function, $\beta^i$ is the shift vector, and $i,j$ run over $r, \theta$ and $z$.
These indices are raised and lowered with the spatial metric $\gamma\ij$.
$\gamma\ij$ is conformally decomposed as $\gm\ij=e^{4\psi}\bar{\gm}\ij$, with the conformal exponent $\psi$ defined as
\begin{align}
    \psi=\frac{1}{12}\ln\left(\frac{\gamma}{\hat{\gamma}}\right),
\end{align}
where  $\hat{\gamma}\ij=\diag (1,r^2,1)$ is the flat reference spatial metric in cylindrical coordinates, and $\gamma$ and $\hat{\gamma}$ are the determinants of $\gamma\ij$ and $\hat{\gamma}\ij$, respectively.
The extrinsic curvature is also decomposed as $K\ij=e^{4\psi}(\bar{A}\ij+\bar{\gm}\ij K/3)$, where $K=K^i_i$ is the mean curvature, and $\bar{A}\ij$ is the conformally decomposed traceless part of $K\ij$.
In the BSSN formalism, a vector quantity $\Delta^i=-\hat{D}_j\bar{\gamma}\IJ$ is introduced as a new independent variable for numerical stability, where $\hat{D}_i$ is the covariant derivative associated with $\hat{\gamma}\ij$.
Due to the cylindrical symmetry, on each slice, only the radial components of $\beta^i$ and $\Delta^i$ are non-zero, while $\gamma\ij$ and $K\ij$ are diagonal.
Consequently, the evolution equations of the BSSN variables $\{\psi,~ \bar{\gamma}\ij,\, K,\, \bar{A}\ij,\, \Delta^r,\, \alpha,\, \beta^r\}$ constitute the gravitational sector in our simulation.

The evolution equations of the BSSN variables with a zero-shift condition, $\beta^i=0$, read 
\begin{align}
\begin{split}
  \pt{t}\psi 
  =& -\frac{1}{6}\alpha K
  \label{eq:evolut_psi}
\end{split}\\
\begin{split}
  \pt{t}\bar{\gamma}_{ij} =& -2\alpha\bar{A}_{ij}   
  \label{eq:evolut_bargamma}
\end{split}\\
\begin{split}
  \pt{t}K =& \alpha\qty(\bar{A}^{ij}\bar{A}_{ij} + \frac{1}{3}K^2) \\
  &+ 4\pi G\alpha\qty(\rho + S) -D_iD^i\alpha
  \label{eq:evolut_traceK}
\end{split}\\
\begin{split}
  \pt{t}\bar{A}_{ij} =&
  \alpha K\bar{A}_{ij} - 2\alpha\bar{A}_{ik}\bar{A}^k_j \\
  &+ \alpha e^{-4\psi}\qty({^{(3)}\calR}_{ij}-8\pi GS_{ij}) \\
  &-\alpha e^{-4\psi}\frac{1}{3}\gamma_{ij}\qty({^{(3)}\calR}-8\pi GS) \\
  &- e^{-4\psi}\qty(D_iD_j\alpha - \frac{1}{3}\gamma_{ij}D_kD^k\alpha)
  \label{eq:evolut_barA}
\end{split}\\
\begin{split}
  \pt{t}\Delta^i =& -2\pt{j}\alpha \bar{A}^{ij} + 2\alpha\biggl[6\bar{A}^{ik}\pt{k}\psi + \bar{\Gamma}^i{}_{jk}\bar{A}^{jk} \\
  &- \bar{\gamma}^{ik}\qty(\frac{2}{3}\pt{k}K + 8\pi GS_k)\biggr]
  ,\label{eq:evolut_Delta}
\end{split}
\end{align}
where $D_i$ is the covariant derivative associated with $\gamma\ij$, the three-dimensional Ricci tensor is given by
\begin{equation}
\begin{aligned}
  {^{(3)}\calR}_{ij} =& {^{(3)}\bar{\calR}}_{ij} + 4\pt{i}\psi\pt{j}\psi - 2\pt{i}\pt{j}\psi + 2\bar{\Gamma}^k{}_{ij}\pt{k}\psi \\
  &- 2\bar{\gamma}_{ij}\qty(2\bar{\gamma}^{kl}\pt{k}\psi\pt{l}\psi + \bar{\gamma}^{kl}\pt{k}\pt{l}\psi - \Delta^k\pt{k}\psi) \\
  {^{(3)}\bar{\calR}}_{ij} =& -\frac{1}{2}\bar{\gamma}^{kl}\pt{k}\pt{l}\bar{\gamma}_{ij} + \frac{1}{2}\bar{\gamma}_{k\{i}\pt{j\}}\Delta^k + \frac{1}{2}\Delta^k\bar{\Gamma}_{\{ij\}k} \\
  &+ \bar{\gamma}^{kl}\bar{\gamma}^{mn}\qty(\bar{\Gamma}_{km\{i}\bar{\Gamma}_{j\}ln} + \bar{\Gamma}_{kmi}\bar{\Gamma}_{lnj}),
\end{aligned}
\end{equation}
the matter variables defined as
\beae{
\rho &= n^{\mu}n^{\nu}T\munu \\
S_i &= -\gamma_i^{\mu}n^{\nu}T\munu \\
S\ij &= \gamma_i^{\mu}\gamma_j^{\nu}T\munu ~,
\label{eq:comp_emtensor}}
are the components of the energy-momentum tensor of the matter fields $T\munu$, $n^\mu$ is the unit normal vector to the slices and $S=\gamma^{ij}S_{ij}$ .

\subsection{Evolution Equations of Matter Fields}\label{sec:eqs}
For the matter sector, the Lagrangian of the Abelian-Higgs model is given in Eq.~\eqref{eq:rescaled_AH_lagrangian}, and the perfect fluid is described by its energy-momentum tensor~\eqref{eq:PF_emtensor}.
Thus, the full energy-momentum tensor is
\begin{equation}
\begin{aligned}
T_{\mu\nu} =& \eta^2\biggl[2\pt{\mu}f\pt{\nu}f + 2n^2f^2\Big(\delta^{\theta}_{\mu}\delta^{\theta}_{\nu} \\ & \qquad  \quad- (\tilde{A}_{\mu}\delta^{\theta}_{\nu} +    \tilde{A}_{\nu}\delta^{\theta}_{\mu}) + \tilde{A}_{\mu}\tilde{A}_{\nu}\Big) \\
&+ g_{\mu\nu}\Big(-\pt{\lambda}f\pt{}^{\lambda}f - \frac{\beta}{2}(f^2 - 1)^2
\\ & \qquad - n^2f^2(g^{\theta\theta}-2\tilde{A}^{\theta} + \tilde{A}_{\lambda}\tilde{A}^{\lambda}) \Big) \\
&+ n^2\qty(g^{\rho\sigma}\tilde{F}_{\mu\rho}\tilde{F}_{\nu\sigma} -\frac{1}{4}g_{\mu\nu}\tilde{F}_{\rho\sigma}\tilde{F}^{\rho\sigma})\biggr] \\
&+ \qty(\rho_{\mathrm{f}}+p_{\mathrm{f}})u_{\mu}u_{\nu} + p_{\mathrm{f}}g\munu.
\label{eq:EMtensor}
\end{aligned}
\end{equation}

Regarding the evolution equations of the matter fields, we derive a set of first-order differential equation with respect to $t$ from the original equations of motion, suitable for numerical implementation.
The evolution equations of the Abelian-Higgs sector are given by
\beae{
\pt{t}f =& \alpha v \\
\pt{t}v =& \alpha\bigl[\gm^{ij}\qty(D_{i}D_{j}f + \pt{i}\ln\alpha\pt{j}f) + K v \\
&- n^2f\qty(\gm^{\theta\theta} - 2\gm^{\theta i}\tilde{A}_i + \gamma\IJ\tilde{A}_i\tilde{A}_j) - \beta (f^3 - f)\bigr] ~.
\label{eq:scalar_eom}}
For the gauge field, the evolution equations read
\beae{
\pt{t}\tilde{A}_i &= -\alpha\gm\ij E^j \\
\pt{t}E^i &= \alpha K E^i + \epsilon^{ijk}\pt{j}\qty(\alpha B_k) - \alpha j^i ~,
\label{eq:gauge_eom}}
where $
B_i = \frac{\pt{r}g}{\sqrt{\gm}}\gamma_{iz}, \,
j^i = 2 f^2 \gm^{i\theta}(1 - g)$, and $\epsilon_{ijk}$ is the Levi-Civita antisymmetric tensor.
Note that the evolution equation for $E^i$ corresponds to the Amp\`{e}re--Maxwell equation, while the other Maxwell equations are automatically satisfied in this setup. 
For the perfect fluid, the evolution equations are as follows\cite{shibata2015Numerical}:
\beae{
\pt{t} \tilde{\rho} &= -\hat{D}_i (\alpha \tilde{S}^i) + \alpha \tilde{S}^{ij}K_{ij} - \gm^{ij} \tilde{S}_i \partial_j \alpha \\
\pt{t} \tilde{S}_i &= -\hat{D}_j (\alpha \tilde{S}_i^j) -\tilde{\rho} \pt{i}\alpha + \frac{1}{2}\alpha \tilde{S}^{jk} \hat{D}_i \gm_{jk} ~,
}
where 
\beae{
\tilde{\rho} &= e^{6\psi}n^{\mu}n^{\nu}T^{\mathrm{fluid}}\munu \\
\tilde{S}_i &= -e^{6\psi}\gamma_i^{\mu}n^{\nu}T^{\mathrm{fluid}}\munu \\
\tilde{S}\ij &= e^{6\psi}\gamma_i^{\mu}\gamma_j^{\nu}T^{\mathrm{fluid}}\munu ~,}
are densitized components of the perfect fluid energy-momentum tensor.

\subsection{Gauge Condition}
The BSSN formalism possesses coordinate gauge freedom, which is specified through $\alpha$ and $\beta^i$.
The system with an expanding string in FLRW has two different time scales, $\eta^{-1}$ for the string and $H^{-1}$ for the asymptotic region.
A space-dependent lapse function is therefore useful, and we choose the following variation of the Bona--Mass\'{o} slicing conditions,
\begin{align}
    \pt{t}\alpha = -2\alpha K_{\infty}
    \begin{cases}
    1 &(t<t_c)\\
    \tanh^4(kr)&(t>t_c)
    \end{cases} ~,
    \label{eq:lapse_condition}
\end{align}
where $K_{\infty}$ denotes the asymptotic value of $K$ in the FLRW region.
The first condition, applied for $t<t_c$, keeps the lapse spatially uniform until the string core develops at time $t_c$.
The value of $t_c$ must be controlled appropriately to avoid the coordinate singularities due to the excessive slice deformation.
For $t>t_c$, the growth in lapse function is suppressed near the center where the string core is located, while the outer regions continue evolving as before. 
To connect these phase smoothly, necessary for numerical stability, we define the smoothing function $s(t) = (\tanh \omega t + 1)/2$ so that the lapse condition becomes
\begin{align}
    \pt{t}\alpha = -2\alpha K_{\infty}\left[1+\left(\tanh^4(kr)-1\right)s(t-t_c)\right]~.
\end{align}
According to this slicing condition, a boundary $r_u$ can be determined outside which the lapse remains almost uniform.
Outside $r_u$, the BSSN terms involving $D_i\alpha$ and $D_iD_j\alpha$ remain negligible throughout the simulations, so that the slicing there is approximately geodesic. 
Our choice of the parameters in the simulation is $t_c=0.3$, $k=10$, and $\omega=50$, and we set $r_u = 10k^{-1}$ explicitly.
The plot of the lapse function and visualization of $r_u$ are shown in Appendix~\ref{sec:lapse_condition}.
Since the lapse increases in time under this condition, it is useful to define the proper time measured by each Eulerian observer as
\begin{align}
    \tau(r) \equiv \int_0^tdt\, \alpha ~.
    \label{eq:proper_time}
\end{align}

In our simulations, we observed that a zero-shift condition generates fewer constraint violations than other tested coordinate choices, and therefore we keep $\beta^r=0$ as our coordinate choice.

Under the slicing conditions, for $r>r_u$ or $r\ll k^{-1}$, Eulerian observers at fixed $r$ corresponds to ``comoving observers'' in the usual sense of FLRW universe.

\subsection{Initial Data} \label{sec:initial_data}
For the initial data of the numerical relativistic system, we adopt the static string solution in a flat space when solving $f(0, r)$ and $g(0,r)$.
The corresponding cosmic string solution is known as the Nielsen--Olesen vortex~\cite{nielsen1973Vortex}, which is characterized by $\beta$ and $n$.
A Nielsen--Olesen vortex is obtained by solving the static equation of motion (imposing $v=0=E^i$ in Eqs.~\eqref{eq:scalar_eom} and~\eqref{eq:gauge_eom}) with boundary conditions~\eqref{eq:NO_boundary} in Minkowski spacetime.
We obtain it numerically with a relaxation method (see Appendix~\ref{sec:relaxation_method}).
In this solution, $f\sim r^n$ and $g\sim r^2$ near the origin, and they approach unity exponentially far from the origin.

As mentioned in Sec.~\ref{sec:string}, we also need a perfect fluid to realize the asymptotically FLRW universe.
To deal with this, we simply add a fluid energy density to achieve a uniform total energy density which is ten times the maximum energy density of the Nielsen--Olesen vortex.
We also impose $u_{\mu}=\alpha\delta^t_{\mu}$, so that the initial fluid momentum vanishes.

The initial data of the BSSN variables are obtained by solving the Hamiltonian and momentum constraints:
\bae{\calH &= {^{(3)}}\calR + K^2 - K_{ij}K^{ij} - 16\pi G\rho = 0 \label{eq:hamiltonian_constraint} \\
\mathcal{M}^i &= D_jK^{ij} - D^iK - 8\pi GS^i = 0,
\label{eq:momentum_constraint}}
where ${^{(3)}}\calR$ is the three-dimensional Ricci scalar.
Although there might be several choices satisfying these requirements, our initial data is computed by imposing
\bege{\gm\ij=\hat{\gm}\ij \\
\bar{A}^i_j = 0~.}
Under these assumptions, the momentum constraint equation~\eqref{eq:momentum_constraint} is identically satisfied, and the Hamiltonian constraint~\eqref{eq:hamiltonian_constraint} reduces to the first Friedmann equation in a flat space $K^2-24\pi G\rho=0$.
Thus, the initial data can be identified as a flat FLRW universe, where the constituents are not only the perfect fluid but also the Nielsen--Olesen vortex.
The initial Hubble parameter $H_{\ini}$ is defined via $3H_{\ini}=-K$.
We choose the initial lapse function to be unity and the shift vector to be zero. 

The remaining parameters required for the simulation are $\eta$, $\beta$ and $n$.
The threshold value of $\eta$ for topological inflation has been reported in~\cite{cho1998Inflation} for a vacuum background, and it is almost the same value in our simulation.
The winding number $n$ determines the behavior of $f(t,r)$ near the origin, $f\sim r^n$, even in curved spacetime.
Thus, $n>1$ displaces the gradient energy slightly away from the origin, which counters the shrinking effect of vacuum energy and helps us to achieve long-time physically stable simulation.
We then choose $\eta=0.3/\sqrt{G}$, $\beta=1$ and $n=2$.

\section{Numerical Result}\label{sec:result}

\subsection{Dynamics of a Self-Gravitating Local String}
\label{sec:string_dynamics}

We now describe the evolution of the matter fields that gives rise to the dynamical string and to the geometrical structures discussed below.  The initial data in our simulations consists of a Nielsen--Olesen vortex embedded in a radiation background, balanced gravitationally by solving the Hamiltonian and momentum constraints as described in Sec.~\ref{sec:initial_data}. The setup begins with a configuration where the scalar winding and the associated magnetic flux are already present on the initial slice. The simulation then follows the subsequent development of the self-gravitating string in a dynamical spacetime. 

In the static case, the vortex is maintained by a balance between the spatial gradients in the field equations and the scalar potential.
In a dynamical spacetime, the static balance is disrupted as the metric and the gradients evolve, forcing the scalar and gauge profiles to readjust.  
Near the core, the electromagnetic and fluid energy densities decrease, while the boundary condition $f(t,0)=0$ keeps the scalar at the potential maximum.
The nonzero potential energy consequently comes to dominate the central region and triggers a sustained de-Sitter like inflationary expansion at the string's core. 
Away from the core, the weakening of gradient support allows the fields to roll toward their vacuum configurations ($f=1$ and $g=1$), and subsequently oscillate about the vacuum values.

This evolution leads to the formation of an inflating core, surrounded by a region dominated by oscillating string fields, followed by an asymptotic radiation-dominated exterior.  The different expansion histories of these regions give rise to a non-monotonic circumferential radius and the associated belly and throat examined below.

\begin{figure*} 
	\centering
	\includegraphics[width=160mm]{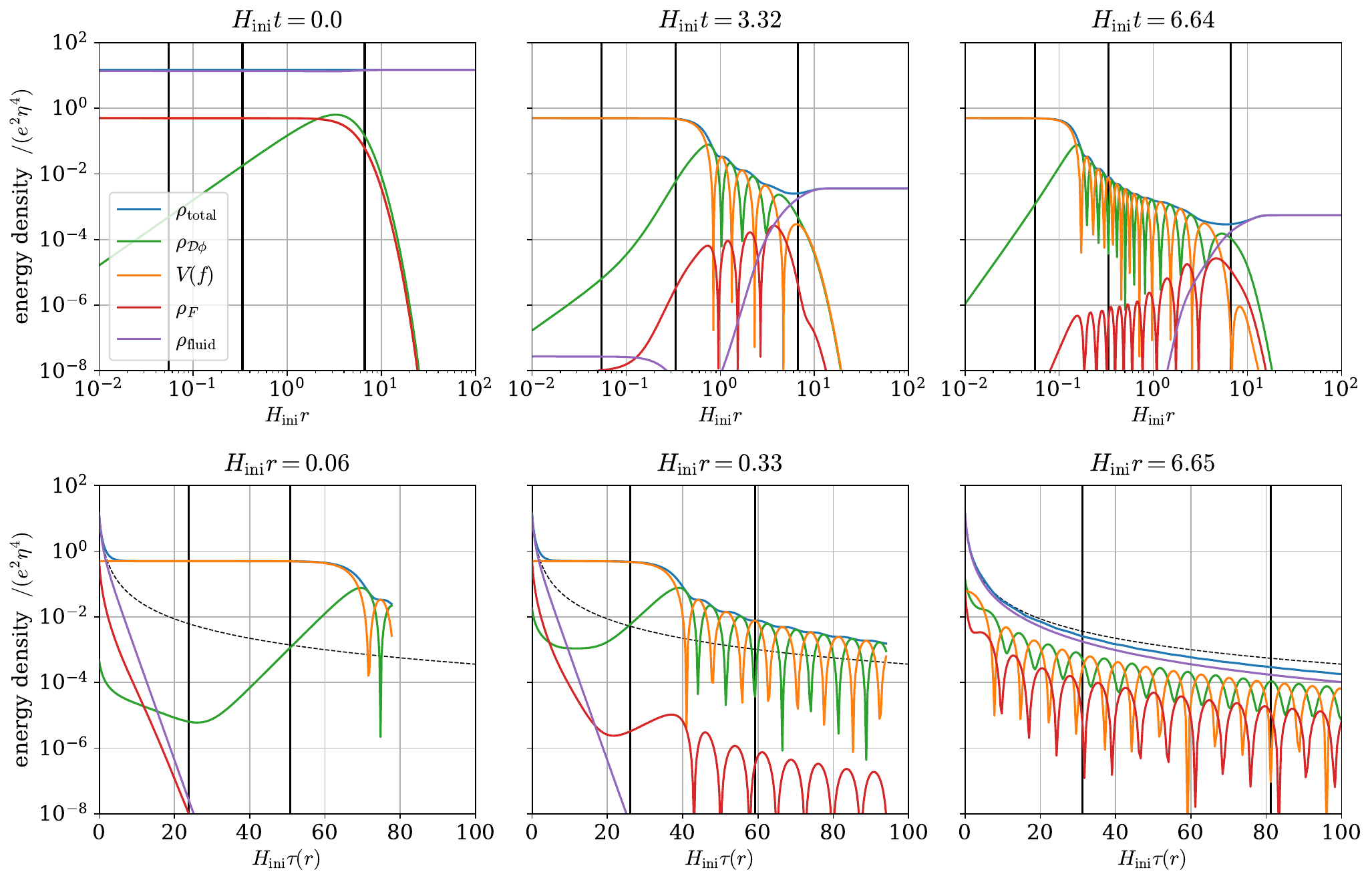}
	\caption{Evolution of the energy densities normalized by $e^2\eta^4$ in each sector. The blue lines represent the total energy density. The orange and green lines denote the covariant derivative terms and potential terms of Abelian-Higgs sector, respectively. The red lines denote the electromagnetic kinetic energy density, and the purple lines correnspond to the energy density of the radiation fluid. The upper panels show the spatial distribution at different times, while the bottom panels show the evolution with respect to $\tau(r)$ at each fixed $r$.
    The vertical black lines in the upper (lower) panels indicate the specific spatial positions (times) plotted in the lower (upper) panels. The black dashed lines in bottom panels are the radiation energy evolution in the asymptotic region. In the top and left panel, the red line and the orange line are overlapped.}
	\label{fig:rho_evol}
\end{figure*}

\begin{figure}[ht]
    \centering
      \includegraphics[width=80mm]{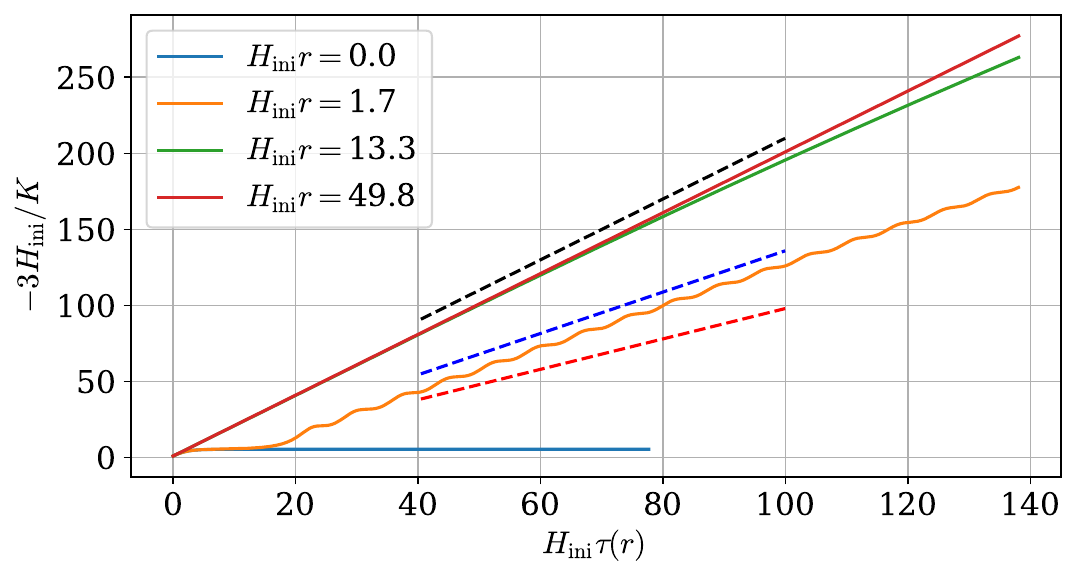}
    \caption{Evolution of the mean curvature with respect to $\tau(r)$ at each fixed $r$. The black and red dashed lines represent  proportionalities of $2H_{\ini}\tau$ and $H_{\ini}\tau$, expected for equation of state parameters $w=\frac{1}{3}$ and $-\frac{1}{3}$, respectively, while the blue dashed line corresponds to $1.36H_{\ini}\tau$, fitted to the orange solid line.} 
    \label{fig:hubble_evol}
\end{figure}

\begin{figure*}[ht]
    \centering
    \includegraphics[width=80mm]{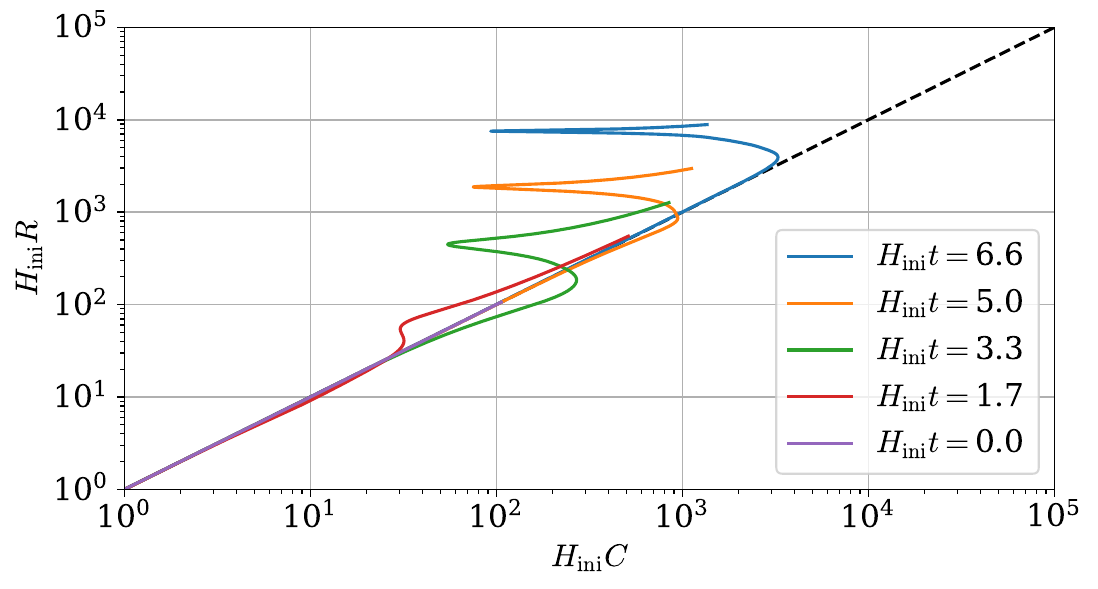}
    \includegraphics[width=80mm]{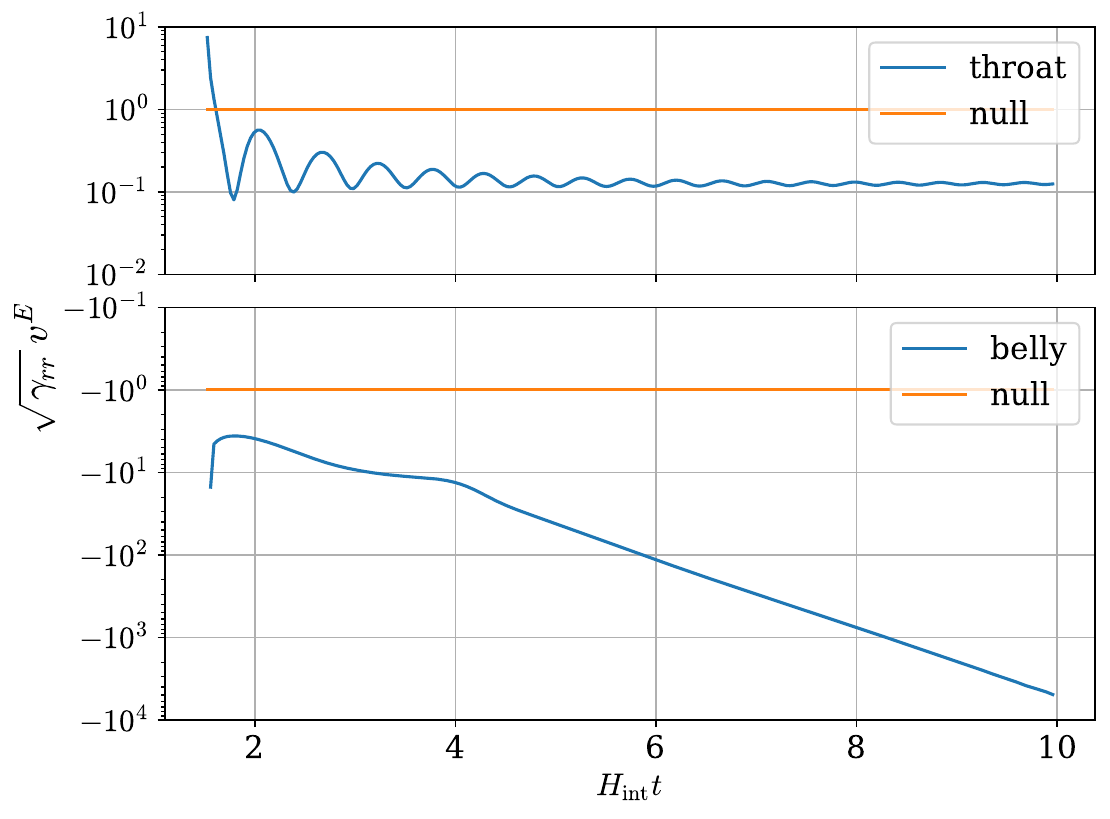}
      
    \caption{Left panel shows the relation between two physical distances measured from the symmetric axis on each slice, defined in Eqs.~\eqref{eq:distances}. The black dashed line denotes the relation in the flat FLRW universe.
    Right panel, the throat (top) and belly (bottom) velocities are represented by the blue lines, while the orange lines correspond to the velocity of null geodesics. These lines are in terms of the proper radial velocity measured by an Eularian observer on each slices: $v^E$.}
    \label{fig:RvsC_throat_velocity_evol}
\end{figure*}

\subsection{Kinematics, Expansion History and Spatial Geometry}

The energy densities of the different matter components, shown in Fig.~\ref{fig:rho_evol}, are defined as
\begin{align}
    \rho_{\mathrm{total}} &\equiv \rho_{\calD\phi}+V(f)+\rho_{F}+\rho_{\mathrm{fluid}} \\
    \rho_{\calD\phi} &\equiv v^2+\pt{i}f\pt{}^if+n^2f^2(1-g)^2\gamma^{\theta\theta} \\
    V(f) &\equiv \frac{\beta}{2}\qty(f^2 - 1)^2 \\
    \rho_{F} &\equiv \frac{n^2}{2}\qty(E^iE_i+B^iB_i) \\
    \rho_{\mathrm{fluid}} &\equiv \qty(W^2\rho_{\mathrm{f}}+ (W^2 - 1)p_{\mathrm{f}})/\eta^2
\end{align}
where $W\equiv -n_{\mu}u^{\mu}$ is the Lorentz factor, and every energy density is normalized by $e^2\eta^4$.
The local volume expansion rate is measured by the mean curvature $K$, with $-K/3$ reducing to the Hubble parameter in an FLRW universe. Fig.~\ref{fig:hubble_evol} shows its inverse, $-3H_{\ini}/K$, at several radius as a function of the proper time.
At $r=0$, the mean curvature becomes constant, which corresponds to de Sitter expansion.
In the matter sector, the expanding core of the string causes the contributions of the electromagnetic fields and the perfect fluid to decay rapidly, leaving the constant vacuum energy (bottom left panel of Fig.~\ref{fig:rho_evol}).
The gradient energy of the scalar field also decays, but it subsequently begins to grow off-center in the core.
After the gradient energy reaches a sufficiently high value, the energy exchange between the vacuum energy and the gradient energy becomes dominant (bottom panels of Fig.~\ref{fig:rho_evol}).
This indicates that the false vacuum core shrinks in the comoving frame due to its negative radial pressure, causing the fields in the Abelian-Higgs sector to fall toward their vacuum configurations and begin to oscillate.
In this region, both the scalar field and the gauge field are massive and they decay at almost the same rate, while the electromagnetic energy remains subdominant (bottom panels of Fig.~\ref{fig:rho_evol}).
The evolution of the mean curvature in this region implies that these oscillating fields act as sources that decelerate the expansion of the volume element.
In particular, the orange line in Fig.~\ref{fig:hubble_evol} exhibits a behavior close to that of a matter-dominated universe, but with a slightly negative pressure (equation of state parameter $w\approx-0.1$), implying that space is not completely flat.
This decelerated expansion connects continuously to the asymptotic region where the dominant source becomes the perfect fluid.
In the asymptotic region, the volume expansion matches that of the radiation-dominated FLRW universe, as expected.
Notably, the expansion rate never falls below that in a radiation-dominated universe throughout the simulation, and there is no contracting feature that could lead to gravitational collapse.

\begin{figure*} 
	\centering
	\includegraphics[width=160mm]{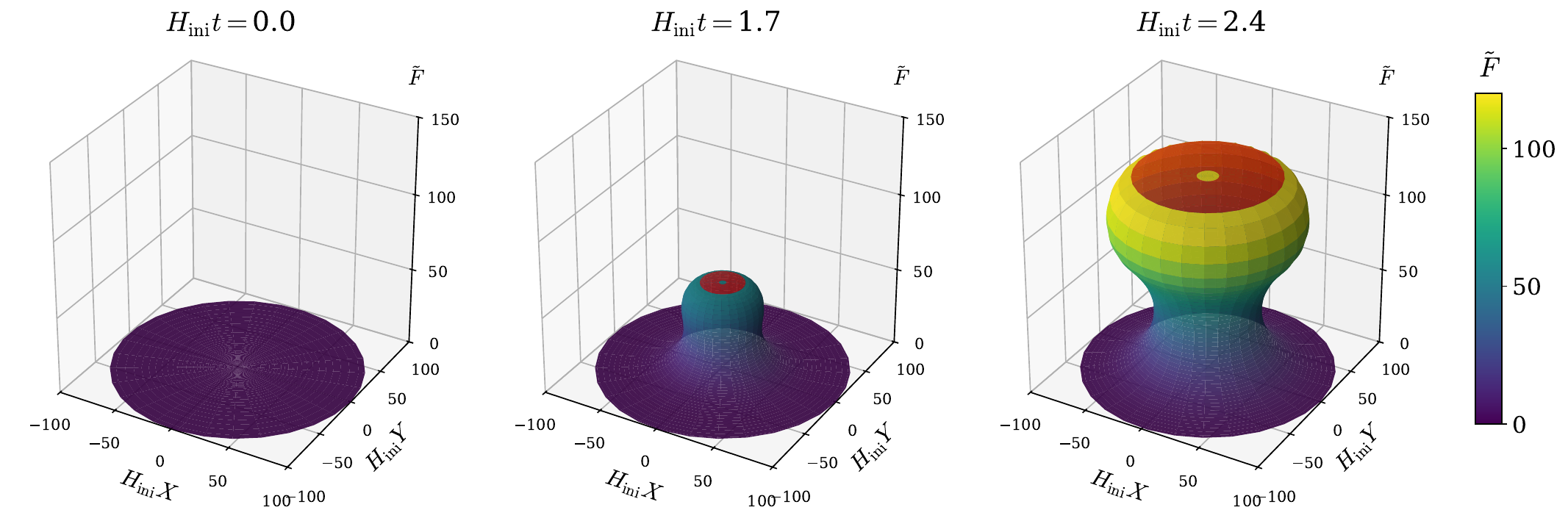}
	\caption{$r\theta$-surfaces embedded in three-dimensional Euclidean space during the early stages of the simulation, where $(X,Y)=(C\cos\theta,C\sin\theta)$. Although the desired embedding function $F$ is defined by $dF^2=dR^2-dC^2$, $dF^2$ can be negative. Thus, the plots are obtained from the simulation data by integrating $d\tilde{F} = \sqrt{\max(dR^2 - dC^2,0)}$, with the region where $dF^2<0$ highlighted in red.}
	\label{fig:embedding}
\end{figure*}

These expansion features are reflected in the left panel of Fig.~\ref{fig:RvsC_throat_velocity_evol}, which represents the relation between the proper radius $R$, and the circumferential radius $C$, defined as follows:
\beae{
R &\equiv \int_0^r dr\,\sqrt{\gamma_{rr}} ~, \\
C &\equiv \sqrt{\gamma_{\theta\theta}}
\label{eq:distances}.}
The left panel of Fig.~\ref{fig:RvsC_throat_velocity_evol} shows that $C$ has a local maximum and a local minimum on 
every slice after $H_{\rm ini}t=1.5$.
This non-monotonic behavior of $C$ with respect to $R$ is due to the spatial difference in the expansion rate; $R$ integrates the space-varying volume elements whereas $C$ solely reflects them.
Each value of $R$ uniquely labels every concentric circle, while $C$ does not.
This fact is intuitively visualized in Fig.~\ref{fig:embedding}, which illustrates the evolution of the $r\theta$-surface by embedding it into a three-dimensional Euclidean space.

To characterize this geometry, we now introduce a new quantity $\Delta$ defined as
\begin{align}
    \Delta(t,r) = \frac{1}{2}\int dx^2\,\sqrt{q}\;{^{(2)}\calR} = \pi\int^r_0 dr\,\sqrt{q}\;{^{(2)}\calR}~,
    \label{eq:def_Delta}
\end{align}
where $q_{AB}$ is the induced metric of the two-dimensional $r\theta$-surface $\sigma$, $^{(2)}\mathcal{R}$ is the scalar curvature of $\sigma$, and integration is performed over $[0,r]$.
Due to the cylindrical symmetry, $\Delta$ can also be written in local form as
\begin{align}
    \frac{\Delta}{2\pi} = 1-\pt{R}C ~,
    \label{eq:local_Delta}
\end{align}
where $\pt{R}=\frac{\pt{r}}{\sqrt{\gm_{rr}}}$ is the partial derivative with respect to $R$.
Note that this local form is derived by imposing regularity at the origin, which is satisfied throughout the simulation.
Geometrically, $\Delta$ measures the rotation angle of a vector tangent to $\sigma$ when parallel transported around a circle of radius $r$.
$\Delta$ is a slice-dependent quantity since $\sigma$ is defined on each slice.
However, for a standard light string, it coincides with the deficit angle far from the string in the static case~\cite{ford1981gravitational}, even in a conformally and homogeneously expanding conical structure, which is locally FLRW but globally anisotropic.
Moreover, $\Delta$ vanishes in a standard spatially flat FLRW universe.

\subsection{Evolution of Belly and Throat}
\begin{figure*}[ht]
    \begin{minipage}[c]{.48\textwidth}
    \centering
    \includegraphics[width=80mm]{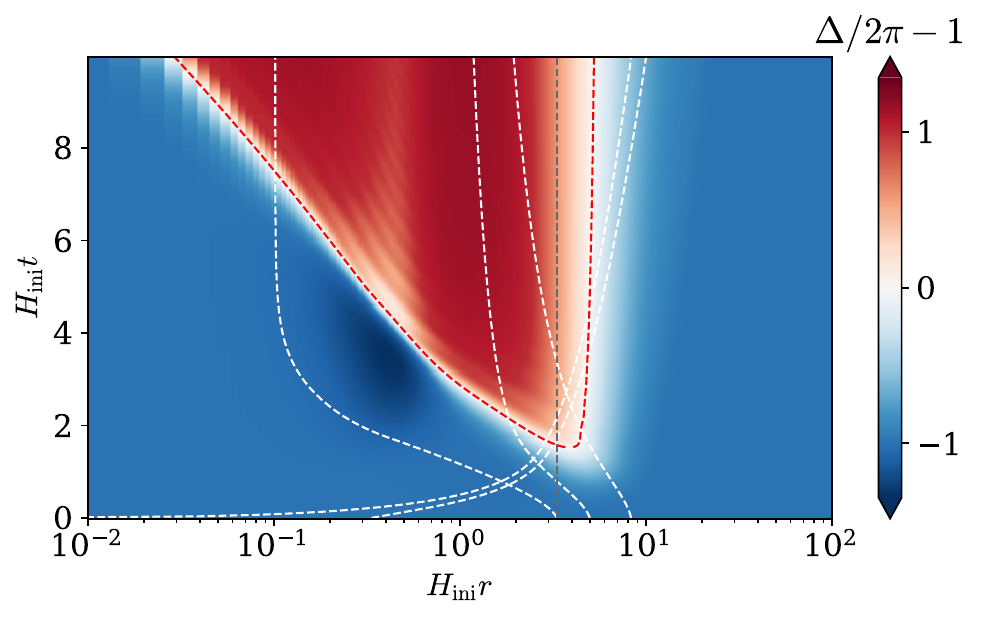}
    \end{minipage}
    \begin{minipage}[c]{.48\textwidth}
    \centering
    \includegraphics[width=80mm]{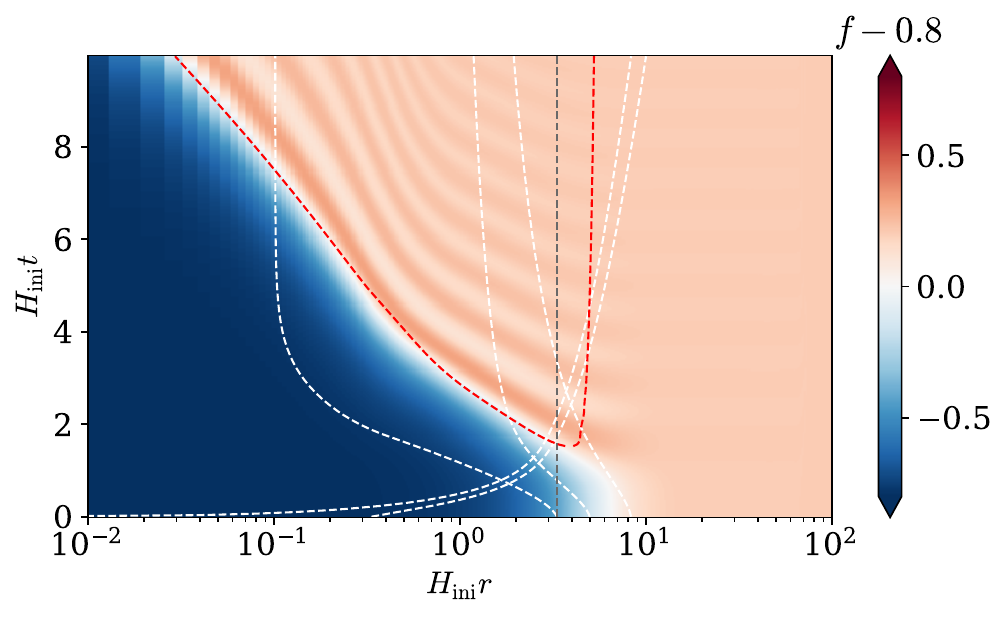}
    \end{minipage}
    \caption{Trajectories of the belly (the left red dashed line) and the throat (the right red dashed line) overlaid on color maps of two quantities. 
    The same red dashed lines, along with white dashed lines (null geodesics) and a vertical gray dashed line ($r=r_u$), are plotted in both panels. The background color plots represent $\Delta/2\pi-1$ (the left panel) and the scalar field configuration $f-0.8$ (the right panel).}
    \label{fig:be_and_th}
\end{figure*}

The embedding diagram in Fig.~\ref{fig:embedding} reveals the development of a belly and a throat.
We define them explicitly as the surface where
\beae{
\mathrm{belly}:&\quad\frac{\Delta}{2\pi} = 1 , \quad \pt{R}\Delta > 0 \\
\mathrm{throat}:&\quad\frac{\Delta}{2\pi} = 1 , \quad\pt{R}\Delta < 0,}
which correspond to the local maximum and minimum of $C$, respectively (see Eq.~\eqref{eq:local_Delta}).
Their trajectories are shown in Fig.~\ref{fig:be_and_th}, overlaid on the evolution of $\Delta$ (left panel) and of the scalar field (right panel).

The belly closely follows the $f=0.8$ contour of the scalar field, which corresponds to the edge of the inflating string core. It moves inwards at a superluminal speed.
This behavior is consistent with that found for other inflating topological defects~\cite{cho1997Spacetime, sakai1996Dynamicsb}.
Because the core is inflating, however, the belly keeps receding from the center in physical distances, and both $R$ and $C$ grow monotonically along its trajectory, as shown in the left panel of Fig.~\ref{fig:RvsC_throat_velocity_evol}.

On the other hand, the throat trajectory is located near the outer edge of the region dominated by the oscillating string fields.
To quantify its motion, we compute the radial Eulerian velocity $v^E = \frac{1}{\alpha}\frac{\dd r(t)}{\dd t}$ (note that the shift vector is zero in our simulation).
Since the throat corresponds to a local minimum of $C$, the radial coordinate velocity of the throat is given by
\bae{\left.\frac{\dd r(t)}{\dd t}\right|_{\mathrm{throat}} = -\left.\frac{\pt{r}\pt{t}C}{\pt{r}^2C}\right|_{\mathrm{throat}}.}
The velocity of the belly is obtained in the same way.
The right panel of Fig.~\ref{fig:RvsC_throat_velocity_evol} shows the resulting physical radial velocities $\sqrt{\gamma_{rr}}\cdot v^E$ of the belly and the throat.
Shortly after the belly and the throat form at around $H_{\rm ini}t \approx 1.5$, the  outward velocity of the throat relaxes to approximately $10\%$ of the speed of light and remains mostly unchanged until the end of the simulation. 
This indicates that the dynamics of the throat differ from those of a point mass, whose peculiar velocity decays due to Hubble friction in an expanding universe.
Since the throat keeps a nearly constant peculiar velocity, this implies that every comoving observer in the outer FLRW universe is eventually overtaken by the throat.

\subsection{Gravitational Wave and Asymptotic Region}
The spatial structure outside the throat can be examined using $\Delta$ and the Weyl invariant $I\equiv C^{\mu\nu\rho\sigma}C_{\mu\nu\rho\sigma}$, where $C^{\mu}{}_{\nu\rho\sigma}$ is the Weyl tensor.
Here, $\Delta$ characterizes the deviation from the flat FLRW universe, while $I$ corresponds to the Weyl curvature, carried by  
gravitational waves and source-free gravitational fields.
For instance, in the conical space generated by a light cosmic string, $I$ vanishes outside the core and $\Delta$ takes a nonzero value, whereas both quantities vanish in a flat FLRW universe.

\begin{figure*}[ht]
    \begin{minipage}[c]{.48\textwidth}
    \centering
    \includegraphics[width=80mm]{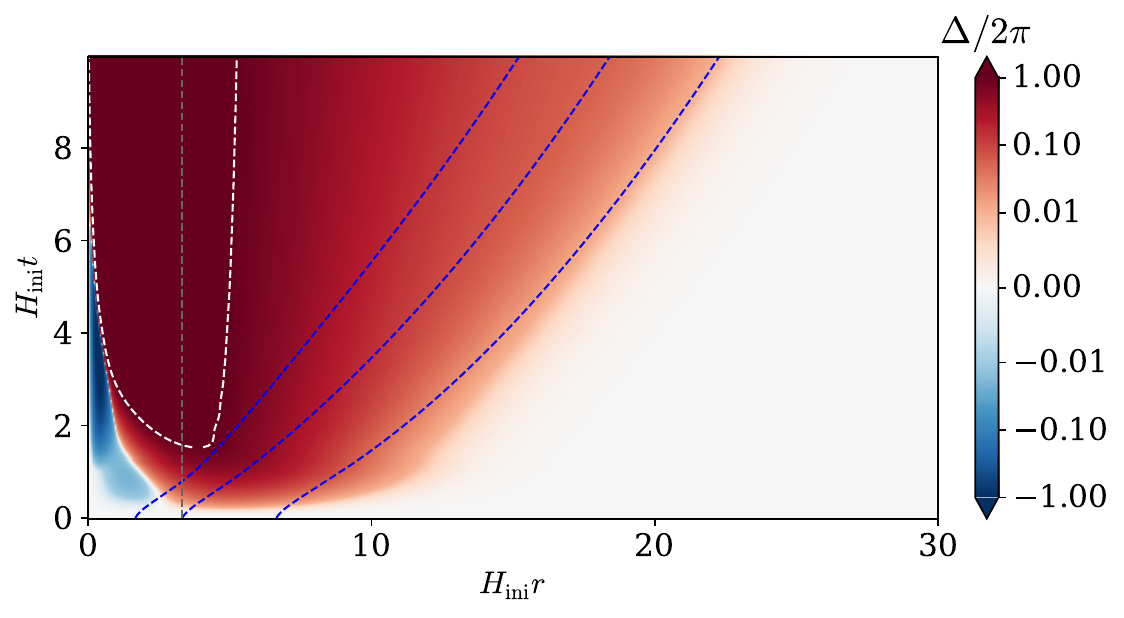}
    \end{minipage}
    \begin{minipage}[c]{.48\textwidth}
    \centering
    \includegraphics[width=80mm]{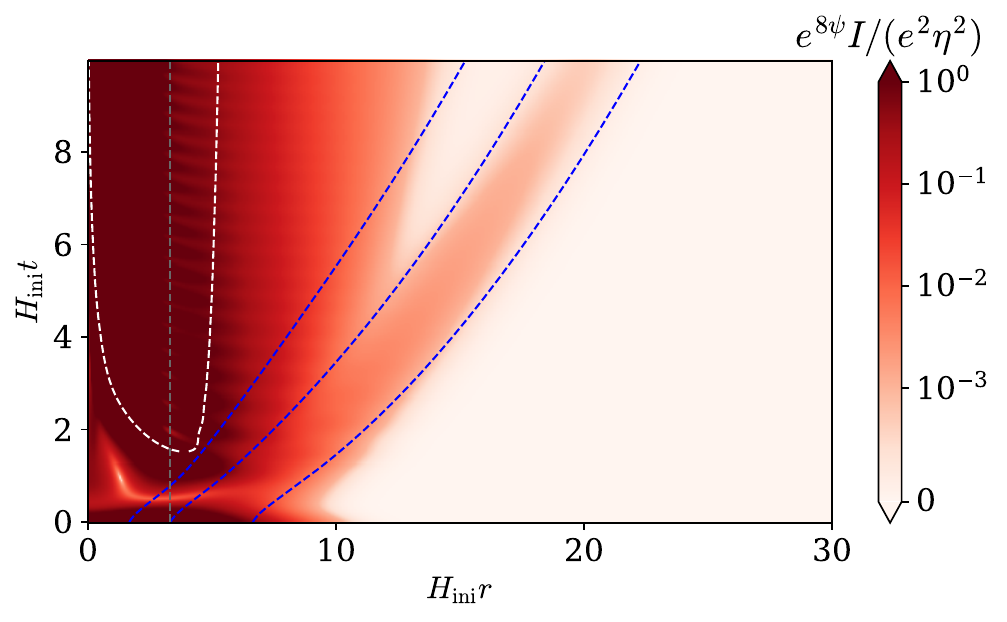}
    \end{minipage}
    \caption{Evolution of $\Delta$ and the scaled Kretchmann scalar $I=C^{\mu\nu\rho\sigma}C_{\mu\nu\rho\sigma}$, where $C^{\mu}{}_{\nu\rho\sigma}$ is the Weyl tensor. The white dashed lines are the trajectory of the belly (left) and the throat (right). The blue dashed lines are outward null geodesics, and the gray dashed vertical line is $r = r_u$.}
    \label{fig:Delta_weyl_color}
\end{figure*}

The evolution of $\Delta$ and $I$ are shown in Fig.~\ref{fig:Delta_weyl_color}.
Far from the inflating string, both vanish as in the FLRW universe.
However, a peak of $I$ is emitted from the initial edge of the string core and propagates at the speed of light, indicating the emission of gravitational wave due to the formation of the string.
This behavior is consistent with the spacetime evolution induced by the formation of a light string in the framework of perturbation theory~\cite{lake2012Evolution}.
Therefore, as required by causality, the nontrivial geometry generated by the string is confined within this wavefront.
Outside the throat, there is a transition region that relaxes into the asymptotic FLRW universe, which spreads at subliminal speed.
Consequently, a region with a non-negligible volume develops between the outer edge of this transition region and the null wavefront, in which $\Delta$ is nonzero.
The right panel of Fig.~\ref{fig:Delta_weyl_color} also shows that the null geodesics initially emitted from the region 
$H_{\ini}r\lesssim3$ 
remains behinds the wavefront.

Fig.~\ref{fig:Delta_on_null} shows that $\Delta$, evaluated along null geodesics that reach a region where $K\approx
-3H$, eventually decays as $e^{-\psi_{\infty}}=a^{-1/2}$, where $a$ is the scale factor of the asymptotic FLRW universe.
Since the expansion of the region where $K\approx -3H$ closely follows that of a radiation-dominated FLRW universe, $a$ grows proportionally to the conformal time, and with the commoving distance $r$.
Thus, $\Delta\propto r^{-\frac{1}{2}}$ along the null geodesics at late times.
Fig.~\ref{fig:Delta_on_null} also indicates that $\Delta$ has no offset anywhere inside the wavefront at future null infinity.
We note that $\Delta$ measured by Eulerian observers is significantly different, as they are eventually overtaken by the throat and enter Region IV, where $\Delta$ is of order unity.

\bfe{width=80mm}{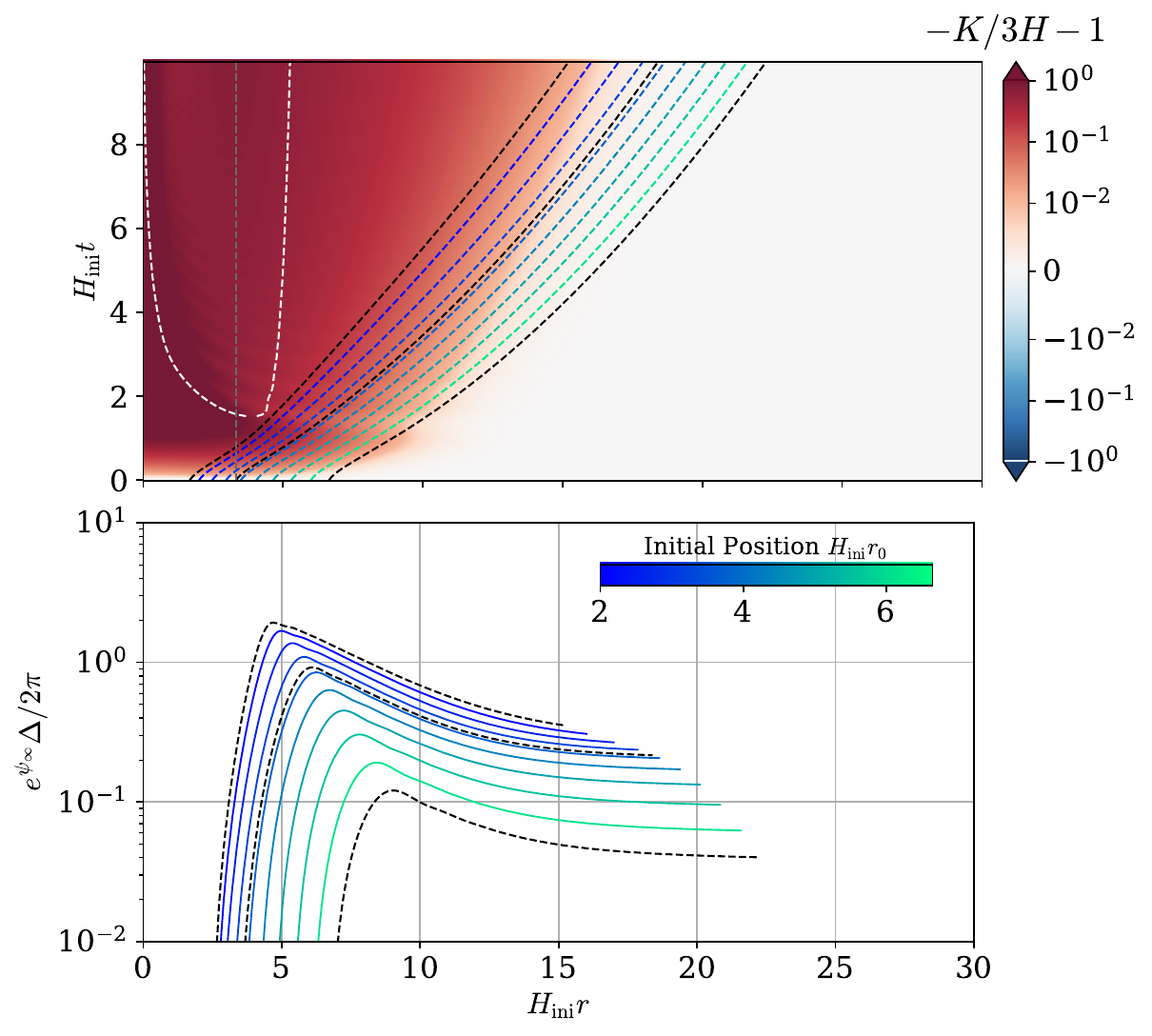}{$-K/3H-1$ and the scaled $\Delta$ measured along null geodesics. Top figure shows the geodesics on a color plot of $-K/3H-1$, where $H$ is the Hubble parameter of the asymptotic FLRW universe. Bottom graph shows plots of $e^{\psi_{\infty}}\Delta/2\pi$ on each corresponding null geodesics, where $e^{\psi_{\infty}}=\sqrt{a}$ is the square root of the scale factor $a$ in asymptotic FLRW universe. The initial position $r_0$ represents the points where the geodesics are initially emitted. The black lines are the same geodesics as those in Fig.~\ref{fig:Delta_weyl_color}. The left and right white dashed lines are the belly and the throat, respectively.}{fig:Delta_on_null}

\section{Discussion and Conclusion}\label{sec:discussion}
\begin{figure} 
	\centering
	\includegraphics[width=80mm]{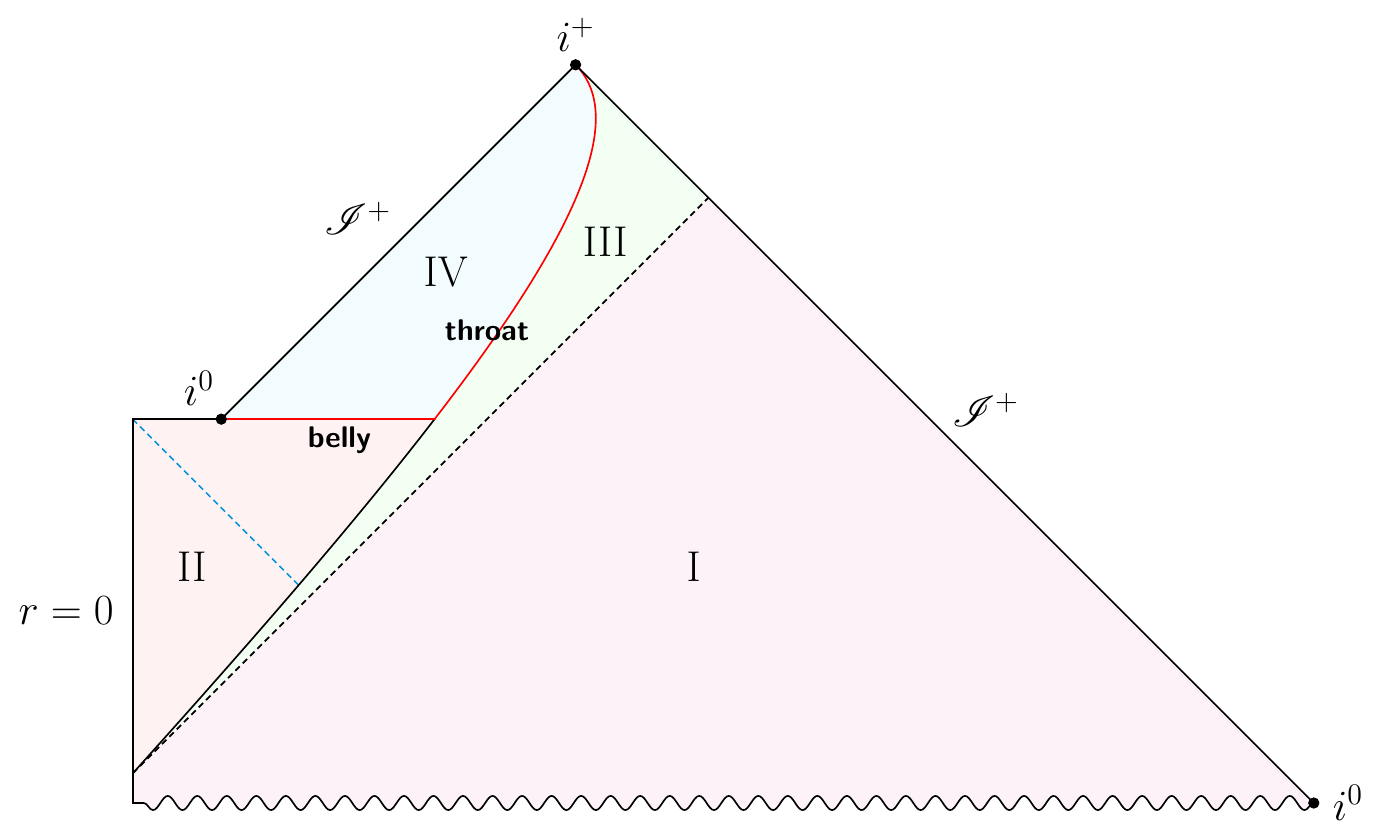}
	\caption{Penrose diagram of spacetime evolution by the massive string in FLRW universe. The black and red lines are the boundary of each regions, while the dashed lines denote null geodesics. $i^+$, $i^0$ and $\mathscr
    I^+$ represent future timelike infinity, spacelike infinity and future null infinity, respectively. Note that the size of each region has no significance since the Penrose diagram just reflects the causal structure.}
	\label{fig:penrose_diag}
\end{figure}

In this paper, we investigated the spacetime evolution induced by the formation of a massive cosmic string using numerical relativity.
We summarize our numerical results as the Penrose diagram pictured in Fig.~\ref{fig:penrose_diag}.
The whole spacetime consists of four regions:
\begin{itemize}
  \item Region I (magenta) is an FLRW universe possessing an initial singularity.
  \item Region II (red) is a de Sitter universe, representing the inflating core of the massive string.
  \item Region III (green) corresponds to the radiation-dominated FLRW universe which the null wavefront has already reached.
  \item Region IV (cyan) is a nearly matter-dominated FLRW universe undergoing decelerated expansion due to the string fields oscillations.
\end{itemize}
The left vertical boundary of the diagram denotes the center of the string core, while the boundary between regions II and IV is identified as the belly, and the one between regions III and IV denotes the throat. 
As is customary, $i^0$, $i^+$ and $\mathscr{I}^+$ denote spacelike infinity, future timelike infinity and future null infinity, respectively. 
Since the throat moves outward at a nearly constant velocity, its trajectory asymptotically approaches future null infinity.
This indicates that any comoving observer in Region III enters Region IV before reaching future timelike infinity.
The upper spacelike edge of Region II reflects the stability and persistency of local string.
The string core (Region II) and the outer region (Region III and IV) are eventually causally disconnected, as the belly moves inward at a superluminal speed.
Furthermore, there is no singularity other than the initial one in the entire spacetime.
This is distinct to other topological inflating scenarios, such as false-vacuum bubbles, which can collapse into black holes~\cite{blau1987Dynamicsb, farhi1990It, sakai2006Universe}.
Solving problems arising from the initial singularity requires other physics, but it is out of our scope.

The superluminal nature of the belly arises from the classical tachyonic instability near the potential maximum.
For a standard cosmic string, the core is stabilized by a balance between the gradient energy and the instability due to the negative mass-squared term.
For an inflating string, however, the rapidly expanding core leads to the dilution of the gradient energy.
As a result, the fields roll down toward the vacuum configuration, mirroring a classical second-order phase transition.
Consequently, in Region IV, the belly can be observed as a surface of a simultaneous second-order phase transition after topological inflation.
This superluminal feature of the belly differs from the dynamics of the shell forming the edge of false vacuum bubbles \cite{farhi1990It, blau1987Dynamicsb, sakai2006Universe}.
The bubble is typically bounded by a shell that is assumed to trace out a timelike world volume.
Since this shell is a physical object carrying concentrated energy density, its propagation cannot exceed the speed of light.
In contrast, the belly in our system is merely an edge of the string core, and its velocity is a phase velocity.

The dominant matter component in Region IV is the oscillating scalar field in the Abelian-Higgs sector, with a subdominant gauge-field contribution.
Thus, Region IV behaves like a matter-dominated FLRW universe, decelerating the volume expansion.
All BSSN variables continuously connect to Regions II and III, exhibiting no rapid changes.
This smooth transition indicates the absence of any thin-shell structures between these regions.
It is worth noting that if the massive string is considered to form during a second-order phase transition as mentioned in the first paragraph of Sec.~\ref{sec:string}, the oscillating string fields in Region IV should decay into radiation fluid via the same reheating-like process that  generated the initial radiation fluid after the transition.
This conversion, however, does not alter the Penrose diagram in Fig.~\ref{fig:penrose_diag}, since the radiation fluid also depicts a decelerating expanding universe.

Finally, the difference between Regions I and III lies in whether the gravitational waves due to the string formation have arrived or not.
In other words, Region III is causally connected to the string formation while Region I is not.
In terms of $\Delta$, while it vanishes in Region I, it can take a non-zero value in Region III.
However, the value of $\Delta$ decreases as $a^{-1/2}$ along outgoing null geodesics, approaching zero far from the throat, or equivalently, at late times after the string formation.
This implies that $\Delta$ becomes negligibly small throughout Region III at late times, and no ``deficit angle'' offset appears on future null infinity.

Our numerical simulation provides a consistent physical picture of the spacetime evolution after the formation of a massive cosmic string.
Furthermore, we have simulated the same system with different initial data, which is obtained by imposing a homogeneous fluid energy density, instead of imposing a homogeneous total energy density while preserving the localization of the string's gravitational effect.
Although this change introduces initially localized inhomogeneity and anisotropy, the additional simulation yields qualitatively identical results compared to the original one.
This suggests that the causal structure and the classical dynamics described above are general consequences of the system.

\hfill

\begin{acknowledgments}
The authors thank Katy Clough and the members of the GRTL collaboration for their contribution to the \texttt{engrenage} code. 
This work is supported by the JSPS KAKENHI Grant Number 26K07083 (TS). 
C.J. is supported by the NSFC Grants No. E414660101, and No. 12547104,
and by the Fundamental Research Funds for the Central Universities under Grants No. E4EQ6604X2 and No. E3ER6601A2.
\end{acknowledgments}

\bibliographystyle{apsrev4-1}
\bibliography{biblio.bib}

\appendix

\section{Computational Details}
Our numerical implementation is based on an open-source numerical relativity code, \texttt{engrenage}~\cite{2026GRTLCollaboration}.
Since it is the BSSN code in spherical symmetry with a real scalar field as a matter field, we have modified it to  impose cylindrical symmetry.
Furthermore, we introduce a gauge field coupled to a scalar field to realize the Abelian-Higgs model given by~\eqref{eq:AH_lagrangian}, and a perfect fluid with a barotropic equation of state to simulate the asymptotic FLRW universe.

Time integration is performed using a fourth-order Runge--Kutta (RK4) method, and spatial derivatives are discretized using a five-point stencil.
For the spatial discretization, we also use a cell-centered and non-uniform grid $r$ in terms of a uniform grid $x$, defined via $r = c \sinh x$.
The outer boundary of the computational domain is $r_{\max}=32$ and is resolved by $N = 2525$ grid points.
Each boundary has three ghost cells.
We apply the Kreiss--Oliger dissipation to suppress high-frequency noise.

\begin{figure}[ht!]
    \centering
      \includegraphics[width=80mm]{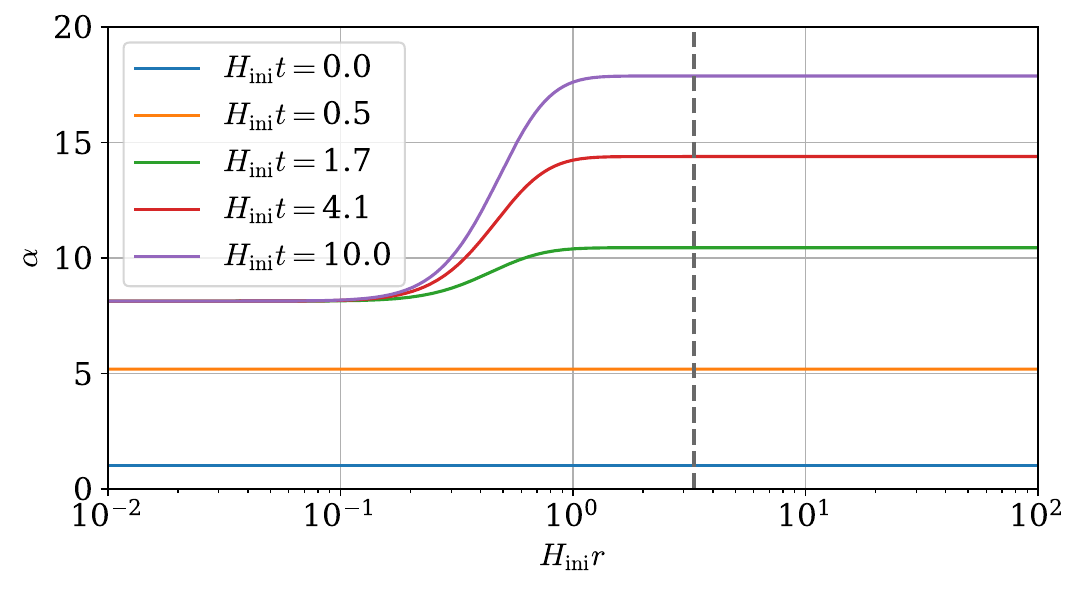}
    \caption{Plot of the lapse function $\alpha$. The black dashed line corresponds to $r_u=10k^{-1}$, beyond which $\alpha$ becomes approximately  flat.}
    \label{fig:lapse_evol}
\end{figure}

\section{Relaxation Method}\label{sec:relaxation_method}
As initial configurations of $f(0,r)$ and $g(0,r)$, we use a Nielsen--Olesen vortex solution in Minkowski spacetime.
To obtain these static configurations, we employ a relaxation method.
Instead of solving the nonlinear elliptic equations directly, we introduce a relaxation parameter $\lambda$ and transform the system into the following evolution equations:
\begin{equation}
\begin{aligned}
    \frac{\pt{} f}{\pt{}\lambda} &= \partial_r^2 f + \frac{1}{r} \partial_r f - \frac{n^2 (1 - g)^2}{r^2} f - \beta f(f^2-1), \\
    \frac{\pt{} g}{\pt{}\lambda} &= \partial_r^2 g - \frac{1}{r} \partial_r g + 2 f^2(1 - g).
    \label{eq:static_eom}
\end{aligned}
\end{equation}
The right-hand sides of Eqs.~\eqref{eq:static_eom} are directly related to the negative variational derivative of the energy per unit length $\mu[f, g]$ in $r\theta$-plane:
\begin{equation}
    \frac{1}{r} \frac{\delta \mu}{\delta f} = - 2 \frac{\pt{} f}{\pt{} \lambda}, \quad 
    \frac{1}{r} \frac{\delta \mu}{\delta g} = - \frac{n^2}{r^2} \frac{\pt{} g}{\pt{} \lambda}.
\end{equation}
By the chain rule, the derivative of $\mu$ with respect to $\lambda$ becomes
\begin{equation}
\begin{aligned}
    \frac{\dd\mu}{\dd\lambda} &= 2\pi \int_0^\infty dr \, r \left( \frac{1}{r}\frac{\delta \mu}{\delta f} \frac{\pt{} f}{\pt{} \lambda} + \frac{1}{r}\frac{\delta \mu}{\delta g} \frac{\pt{} g}{\pt{} \lambda} \right) \\
    &= - 2\pi \int_0^\infty dr \, r \left[ 2 \left( \frac{\pt{} f}{\pt{} \lambda} \right)^2 + \frac{n^2}{r^2} \left( \frac{\pt{} g}{\pt{} \lambda} \right)^2 \right].
\end{aligned}
\end{equation}
This indicates that the system monotonically dissipates energy ($\dd\mu/\dd\lambda\le 0$) during the $\lambda$ evolution, ensuring stable relaxation toward stationary states ($\dd\mu/\dd\lambda=0$).

\begin{figure*}[ht]
    \centering
    \includegraphics[width=160mm]{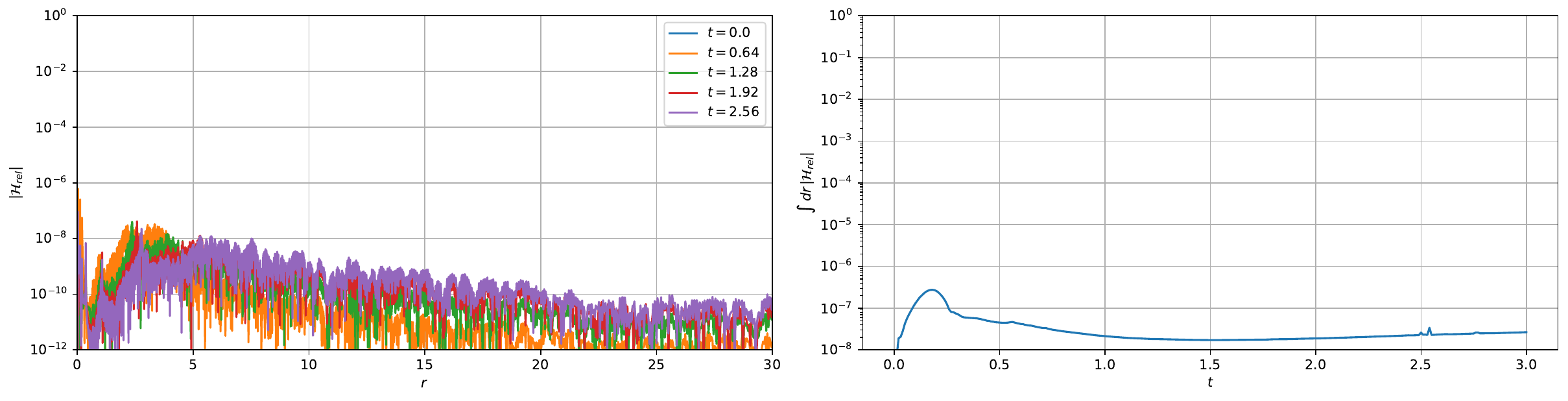}
    \caption{Absolute value of the relative Hamiltonian constraint violation $\abs{\calH_{\mathrm{rel}}}$, defined in Eq.~\eqref{eq:relative_Ham}. The left panel shows the spatial distribution and time evolution of it, while the right panel exhibits the evolution of evolution of the spatially integrated violation $\int dr\,\abs{\calH_{\mathrm{rel}}}$.}
    \label{fig:constraint_converge}
\end{figure*}

Starting from a set of initial guesses of field configurations, the system is evolved in $\lambda$ using RK4 subject to the boundary conditions~\eqref{eq:NO_boundary}.
The relaxation process is terminated when the fields reach a stationary state, defined by the condition that the maximum field updates become sufficiently small values relative to the step size in $\lambda$.

\section{Lapse Condition}\label{sec:lapse_condition}
We impose the lapse condition given by Eq.~\eqref{eq:lapse_condition}.
The resulting evolution of the lapse function $\alpha$ is shown in Fig.~\ref{fig:lapse_evol}.
This plot indicates that a slower evolution takes place near the origin while a faster evolution is realized in the outer region.
Furthermore, the deformation of the lapse function becomes negligible outside $r_u$, where the slices approximately obey the Gaussian slicing condition.

\section{Code Validation}\label{sec:validation}
We check the validity of our simulation by evaluating the relative Hamiltonian constraint violation:
\bae{\calH_{\mathrm{rel}} = \frac{{^{(3)}}\calR + K^2 - K_{ij}K^{ij} - 16\pi G\rho}{\abs{{^{(3)}}\calR} + \abs{K^2} + \abs{K_{ij}K^{ij}} + \abs{16\pi G\rho}} ~,
\label{eq:relative_Ham}}
at each time step.
Fig.~\ref{fig:constraint_converge} shows the code validation of our simulation, and $\abs{\calH_{\mathrm{rel}}}$ is smaller than $10^{-4}$ throughout the simulation.

\end{document}